# Coupling of hole double quantum dot in planar germanium to a microwave cavity

Yuan Kang[1,2,†], Zong-Hu Li[1,2,†], Zhen-Zhen Kong[3,4], Fang-Ge Li[1,2], Tian-Yue Hao[1,2], Ze-Cheng Wei[1,2], Song-Yan Deng[1,2], Bao-Chuan Wang[1,2], Hai-Ou Li[1,2,5], Gui-Lei Wang[3,5,6,*], Guang-Can Guo[1,2,5,*], Gang Cao[1,2,5,*], and Guo-Ping Guo[1,2,5,7*]

1. CAS Key Laboratory of Quantum Information, University of Science and Technology of China, Hefei, Anhui 230026, China
2. CAS Center for Excellence in Quantum Information and Quantum Physics, University of Science and Technology of China, Hefei, Anhui 230026, China
3. Key Laboratory of Microelectronics Devices & Integrated Technology, Institute of Microelectronics, Chinese Academy of Sciences, Beijing 100029, China
4. Institute of Microelectronics, University of Chinese Academy of Sciences, Beijing 100049, China
5. Hefei National Laboratory, University of Science and Technology of China, Hefei, Anhui 230088, China
6. Beijing Superstring Academy of Memory Technology, Beijing 100176, China
7. Origin Quantum Computing Company Limited, Hefei, Anhui 230088, China

† The authors have equally contributed in this article.

*Corresponding author: gcao@ustc.edu.cn, guilei.wang@bjsamt.org.cn, and gpguo@ustc.edu.cn

## ABSTRACT

In recent years, notable progress has been made in the study of hole qubits in planar germanium, and circuit quantum electrodynamics (circuit QED) has emerged as a promising approach for achieving long-range coupling and scaling up of qubits. Here, we demonstrate the coupling between holes in a planar germanium double quantum dot (DQD) and photons in a microwave cavity. Specifically, a real-time calibrated virtual gate method is developed to characterize this hybrid system, which in turn allows us to determine the typical parameters sequentially through single-parameter fitting instead of conventional multi-parameter fitting with additional uncertainty, and gives the hole-photon coupling rate of $g_0/2\pi = 21.7$ MHz. This work is a step toward further research on hole-photon interactions and long-range qubit coupling in planar germanium. The experimental method developed in this work contributes to the more accurate and efficient characterization of hybrid cavity-QED systems.

Holes in semiconductor have become one of the most promising platforms for preparing qubits due to their strong spin-orbit coupling interactions[1-4] and weak hyperfine interactions[5-7]. In germanium, these properties become even more prominent. Its smaller effective mass[8] and stronger Fermi level pinning effect[9] have reduced the complexity and difficulty of the manufacturing process. In recent years, a series of noteworthy experiments related to hole qubit have been demonstrated on germanium platforms[10-15] including qubit manipulation with high fidelity[16]. In terms of qubit scaling, researchers have implemented up to four hole spin qubits in planar germanium[17] and proposed a strategy for jointly controlled electrodes to enable further qubit scaling up[18].

Circuit QED has opened up a pathway for achieving long-range qubit coupling and scalable quantum information processing[19-23]. Significant progress has been made in electron qubits, which have achieved strong coupling between qubits and cavities[24-27] and also the long-range coupling between qubits[28-32]. Regarding hole qubits, substantial advancements have predominantly focused on the nanowire platform. Coupling between cavity photons and holes in quantum dots has been achieved in Ge hut wires[33], as well as strong coupling in etched silicon nanowire structures[34]. However, in planar germanium hole systems, experiments on the coupling between cavity photons and holes are still waiting for exploration.

In this work, we fabricate a hole DQD-cavity hybrid device on planar germanium, and demonstrate the hole-photon coupling. A virtual gate method is developed to characterize the coupling system, which does not require matrix mapping and can be calibrated in real time. This method allows us to obtain the coupling rate and the qubit decoherence rate sequentially through single-parameter fitting rather than typical multi-parameter fitting, which has a larger parameter space and may lead to multiple solutions. By further comparing the extracted qubit decoherence rates from two-tone spectroscopy and cavity transmission spectrum, respectively, the reliability of this method is verified.

As shown in Figure 1a, a high-impedance superconducting transmission cavity made of 10 nm TiN film is connected to the DQD near the antinode of its half-wavelength ($\lambda/2$) mode. The center conductor of the cavity has a length of ~400 μm and a width of ~250 nm. Figure 1b shows the transmission signal of the cavity, without setting up the DQD electrode voltage. Due to the severe Fano effect, the transmission signal becomes a dip instead of a peak[35]. We introduce this effect by incorporating a Fano factor in the input-output theory to fit the transmission signal[36], as shown by the blue line in Figure 1b, and obtain a resonant frequency $f_r = 5.434$ GHz, photon loss rate $\kappa/2\pi = 5.3$ MHz and characteristic impedance $Z_r = 3.0$ kΩ for the cavity.

The structure of the DQD is shown in Figure 1c. The bottom electrodes (SU, SD, and CP) are insulated from the top electrodes (LL, LR, B1, P1, B2, P2, and B3) by a 20 nm $Al_2O_3$ layer. Under the control of these electrodes, holes are trapped in two dots indicated by the white circles in Figure 1c, denoted as (M,N), where M and N represent the number of holes in the left dot and right dot, respectively. A negative voltage of $-0.2$ V is applied to the electrode SD to assist in the opening of the channel and the formation of DQD. The electrode CP is connected to the cavity. Figure 1d shows a cross-sectional schematic of the device along the direction of the P2-CP electrode. More details of the fabrication process are provided in the Supplemental Material.

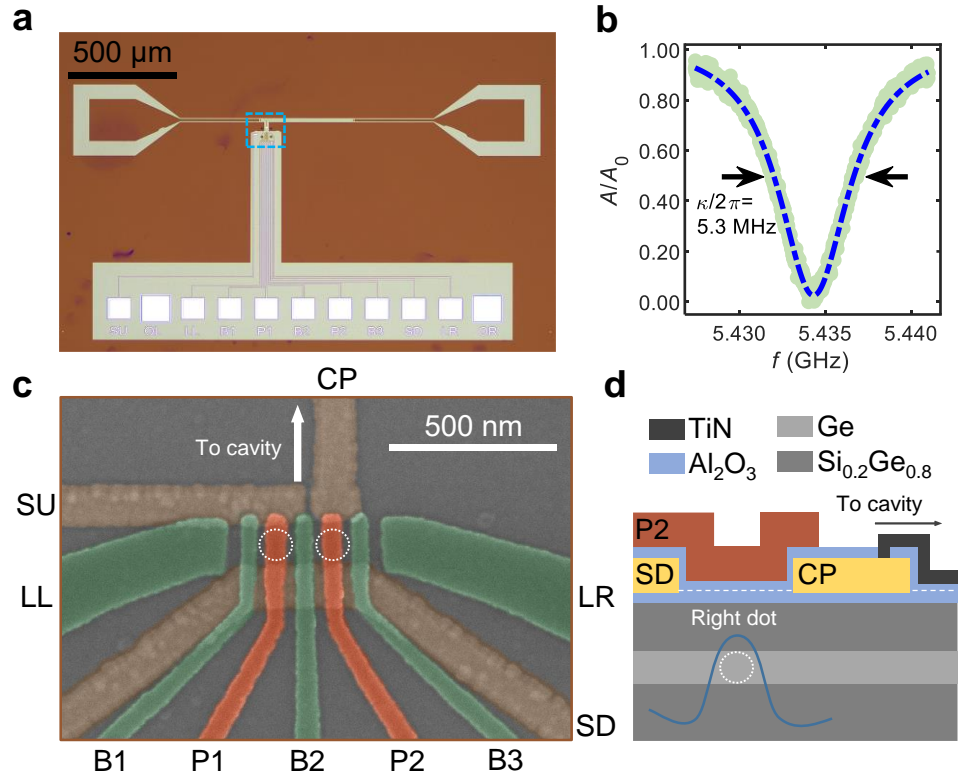

**Figure 1**. Device layout. (a) Optical image of a device similar to the one used in this study, with the blue dashed rectangle indicating the region of the double quantum dot structure. (b) Transmission signal of the cavity (green dots), which performs as a dip rather than a peak due to the severe Fano effect. The blue dashed line illustrates the fitting result after correction with the Fano factor. (c) False-color scanning electron microscope image of the electrodes used for confining holes in quantum dots, with the white circle indicating the formation positions of the quantum dots. (d) Cross-sectional schematic of the device along the direction of the P2-CP electrode, illustrating how the CP electrode is connected to the TiN film through a via. The white dashed line represents the interface between the first and second layers of $Al_2O_3$.

The charge stability diagram of the double quantum dot is shown in Figure 2a, which illustrates a set of charging lines from the right (left) dot to the hole reservoir below LR (LL). We select a working area to characterize the coupling rate, indicated by the red rectangle in Figure 2a, where the hole distribution in the DQD varies between (3,2) and (2,3), and the interdot tunneling signal is shown in Figure 2b.

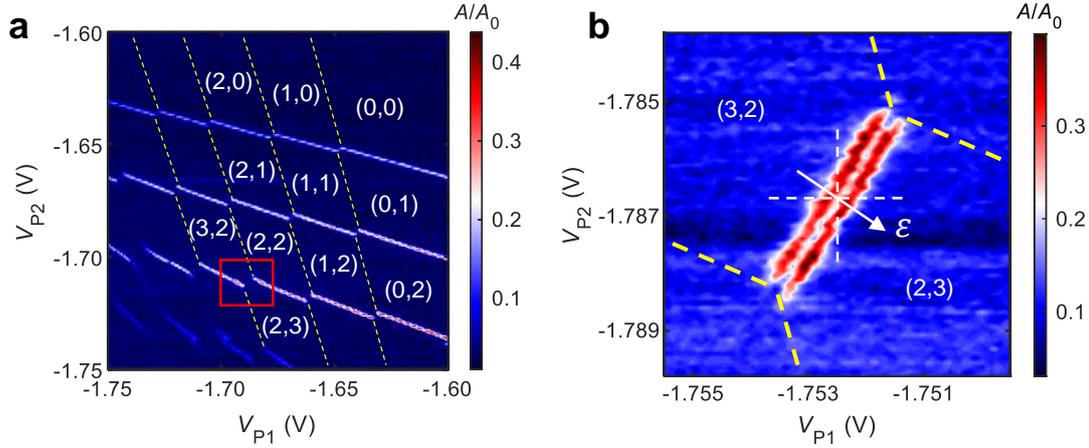

**Figure 2**. Charge stability diagrams of the DQD. (a) The cavity transmission signal $A/A_0$ as a function of voltages on P1 and P2 with $V_{B2} = -1.20$ V. The yellow dashed lines indicates the charging lines from the DQD to the hole reservoir. The region marked by the red rectangle is selected to characterize the coupling rate. (b) Interdot charge transition signal at the point corresponding to the working region selected in (a) with $V_{B2} = -1.10$ V.

Generally, one can characterize the charge-photon coupling rate by measuring vacuum Rabi splitting at the resonance condition[24], where $2t_c/h = f_r$ and $\varepsilon = 0$. Here, $t_c$ is the interdot tunneling coupling rate of the DQD, and $\varepsilon$ is the energy level detuning. However, the splitting could be indistinct under the condition of weak coupling, where the coupling rate is small and the qubit has a large decoherence. In this scenario, it is conventional to use input-output theory to perform multiparameter fitting for the cavity signal influenced by the quantum dot[37, 38]. Specifically, denoting that the decoherence rate of the DQD is $\gamma$ and the resonant cavity is probed at a frequency $f_{\text{probe}}$, coupled to the DQD with coupling strength $g_0$, the cavity transmission signal satisfies[39]:

$$A/A_0 = \frac{-i\sqrt{\kappa_1 \kappa_2}}{\Delta - i\kappa/2 + \widetilde{g_c}\chi}, \qquad (1)$$

where $\Delta = 2\pi(f_{\text{probe}} - f_r)$, effective coupling rate $\widetilde{g_c} = g_0 t_c/hf_a$, dipole moment $\chi = \widetilde{g_c}/[2\pi(f_a - f_{\text{probe}}) + i\gamma/2]$, and qubit transition frequency $f_a = \sqrt{\varepsilon^2 + 4t_c^2}/h$. To extract $g_0$ from the transmission signal, one needs to simultaneously fit $t_c$, $\varepsilon$ and $\gamma$, which is challenging due to the large parameter space and multiple solutions. Although we can calibrate $\varepsilon$ in advance through the lever arm of the electrode and estimate $t_c$ and $\gamma$ through two-tone spectroscopy[40, 24, 25], the former may introduce extra errors, and the latter could be difficult to observe in the weak-coupling regime.

Here, we develop a method to better extract the parameters of the hybrid system. As will be apparent later, although the vacuum Rabi splitting in the resonant regime ($f_a \approx f_r$) is indistinct in our device due to qubit decoherence, a significant cavity frequency shift $\Delta f_r$ is still observable in the dispersive regime ($|f_a - f_r| \gg 10g_0/2\pi$), in which $\Delta f_r$ satisfies[41]:

$$\Delta f_r = \frac{-g_0^2/2\pi}{(f_a - f_r)} \xRightarrow{\varepsilon=0} \frac{-g_0^2/2\pi}{(2t_c/h - f_r)} = -\frac{g_0^2}{2\pi}\delta_c, \qquad (2)$$

where $\delta_c \equiv 1/(2t_c/h - f_r)$. Hence, $g_0$ could be extracted through a single-parameter fitting of $\Delta f_r$ as a function of $2t_c/h$. By gradually adjusting the voltage applied to the B2 electrode, denoted as $V_{B2}$, $2t_c/h$ can be modulated, while $\varepsilon$ is also changed. To independently tune the interdot tunneling, we implement a virtual gate[42-45], denoted as $V'_{B2}$. Here $V'_{B2}$ does not rely on conventional matrix mapping and is calibrated in real-time in the following steps: After each step $V'_{B2}$ changes, $V_{B2}$ is set to $V'_{B2}$, and a fast coarse scan and image recognition are employed in a region similar to Figure 2b. This process fixes $V_{P1}$ and $V_{P2}$ at the center of the interdot charge transition line, as denoted by the white dashed cross in Figure 2b, and ensures that the DQD energy level detuning $\varepsilon$ remains zero during the process of adjusting $2t_c/h$.

The relationship between virtual gate $V'_{B2}$ and $2t_c/h$ is calibrated through two-tone spectroscopy[46]. The probing frequency on the cavity is kept at $f_{probe} = f_r$, and a driving microwave with frequency $f_{pump}$ is applied to electrode P1. Figure 3a shows the transmission signal as a function of $f_{pump}$ and $\Delta V'_{B2}$, where $\Delta V'_{B2} = V'_{B2} + 1.11$ V. The bottom of the amplitude response corresponds to the resonance condition $f_{pump} = f_a$. Since detuning $\varepsilon$ is fixed to 0 during the $V'_{B2}$ tuning process, this response directly corresponds to $f_{pump} = 2t_c/h$, as indicated by the inset in Figure 3a. As the control of $V'_{B2}$ on $2t_c$ is approximately linear in this region, we obtain $\Delta V'_{B2} = -(2t_c/h) \times 2.97$ mV/GHz $+ 13.78$ mV. Besides, we can also roughly estimate the decoherence rate of the qubit from the linewidth of the signal. By fitting the full width at half maximum (FWHM) $\gamma/\pi$ at each $\Delta V'_{B2}$, we obtain an average value of $\gamma/2\pi = 265$ MHz, and Figure 3b depicts three of them as illustrative examples.

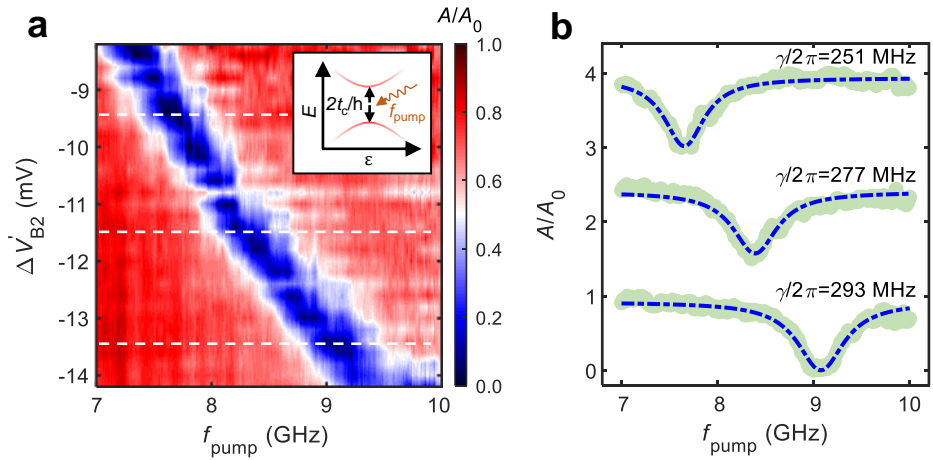

**Figure 3**. Calibrated $V'_{B2}$ through two-tone spectroscopy. (a) The transmission signal of the cavity as a function of $f_{pump}$ and $\Delta V'_{B2}$ at $f_{probe} = f_r$, while $\varepsilon$ is fixed to zero. Inset: Schematic of the energy levels of the DQD. (b) The response of the cavity transmission signal

as a function of $f_{\text{pump}}$ at $\Delta V'_{\text{B2}} = -9.4, -11.6, -13.4$ mV in (a), with each line shifted upward by 1.5. The corresponding quantum dot decoherence rate is obtained through Lorentzian fit.

Figure 4a shows the transmission spectrum of the cavity at $2t_c/h \approx f_r$ (marked with a yellow pentagram). As mentioned above, the vacuum Rabi splitting is indistinct in our device suffering from large qubit decoherence. To extract $g_0$, we continuously vary $2t_c/h$ using virtual gate $V'_{\text{B2}}$ and measure the cavity transmission signal at $\varepsilon = 0$, as shown in Figure 4b. When $2t_c/h < f_r$, as indicated by the yellow triangles in Figure 4a and 4b, we have $\delta_c < 0$ and $\Delta f_r > 0$ in Equation 2. As $2t_c$ increases and approaches $f_r$, $\Delta f_r$ increases accordingly. At $2t_c/h \approx f_r$, the cavity resonant frequency remains unshifted, while the cavity experiences a stronger influence from qubit decoherence, resulting in a larger κ.

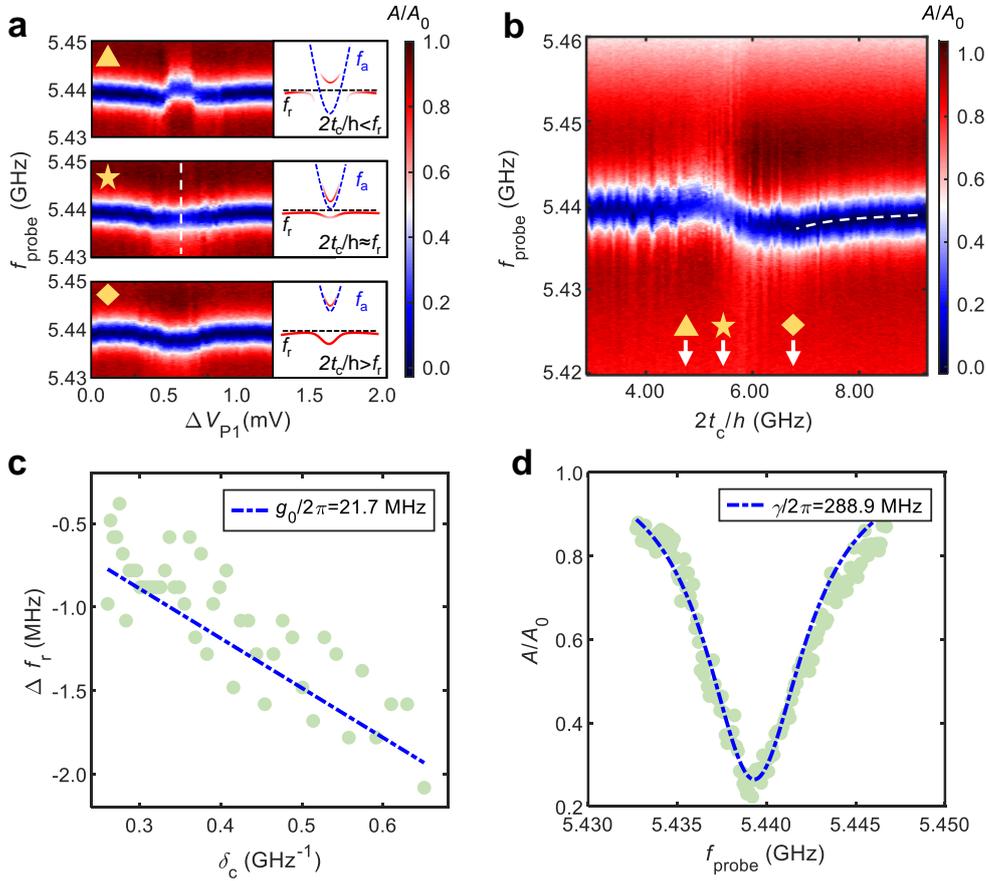

**Figure 4**. Single-parameter fitting of charge-photon coupling rate and qubit decoherence rate. (a) The transmission spectrum of the cavity as a function of $\Delta V_{\text{P1}}$ at $2t_c/h = 4.77, 5.44,$ and $6.79$ GHz, marked by the triangle, pentagram, and diamond, respectively. Insets: Schematics of the relationship between qubit energy and cavity frequency correspondingly. (b) The transmission spectrum of the cavity at $\varepsilon = 0$ changes with a variation in $2t_c/h$. (c) $\Delta f_r$ as a function of $\delta_c$ extracted from the dispersive regime in (b), indicated by the white dashed line. The coupling rate $g_0$ is obtained by linear fitting, depicted

by the blue dashed line. (d) The transmission signal of the cavity at the position indicated by the white dashed line in (a), where $\varepsilon = 0$ and $2t_c/h=f_r$.

When $2t_c/h > f_r$, as indicated by the yellow diamonds, $\delta_c > 0$ and $\Delta f_r < 0$, resulting in a negative shift in the resonant frequency. As $2t_c$ further increases, $\Delta f_r$ gradually returns to 0. The white dashed line in Figure 4b indicates the resonant frequency shift $\Delta f_r$ in the dispersive coupling regime, which can be described by Equation 2. We extract corresponding data to plot the cavity frequency shift $\Delta f_r$ at different $\delta_c$ in Figure 4c and then perform a linear fit. The slope of this line (blue dashed) represents $g_0^2/2\pi$, consequently, we obtain the charge-photon coupling rate as $g_0/2\pi = 21.7$ MHz.

To verify this result, we substitute the obtained $g_0$ into Equation 1 and fit the qubit decoherence rate $\gamma$. Figure 4d shows the transmission signal of the cavity (green dots) at the position indicated by the white dashed line in Figure 4a, where $\varepsilon = 0$ and $2t_c/h = f_r$. The fitting results obtained from Equation 1 give that $\gamma/2\pi = 288.9$ MHz, which is closely consistent with the estimation in Figure 3.

The coupling rate $g_0$ is relatively small, particularly considering that our cavity has a characteristic impedance of up to $Z_r = 3.0$ kΩ, and $g_0 \propto \Delta\alpha f_r\sqrt{Z_r}$[47, 48, 25]. Based on our simulation analysis (see Supplemental Material for more details), we attribute this to the small lever arm of the CP electrode, as well as the impedance mismatch caused by an inappropriate coupling electrode structure which results in reflection of the microwave electric field from the cavity. These effects collectively weaken the effective cavity electric field that reaches the quantum dot, leading to the observed lower coupling rate. On the other hand, the decoherence rate of our qubit is relatively large. This may arise from manufacturing process issues, which results in a leakage from left dot to reservoir due to the limited control ability of electrode B1.

In conclusion, we have demonstrated coupling between hole double quantum dot and a high-impedance cavity in planar germanium. To characterize the hybrid system, a real-time calibrated virtual gate method is developed. This method helps accurately extract coupled system parameters sequentially through single-parameter fitting, avoiding multiple-solution problems caused by multi-parameter fitting. With this method, we obtain a charge-photon coupling rate of 21.7 MHz and the qubit decoherence rate of 288.9 MHz. Our results provide a groundwork for future studies on hole-photon coupling and long-range interactions of hole qubits. With manufacturing and structure improvements in future work, strong coupling is promising in this planar germanium hole-photon coupling system.

## ASSOCIATED CONTENT

**Supporting Information**

Fabrication of the hole DQD-cavity hybrid device (part I), correction of Fano effect (part II), explanation of cavity resonant frequency shift in Equation 2 (part III), simulation of lever arm (part IV), and explanation of the tunneling-depended two-tone spectroscopy in Figure 3a (part V) (PDF)

## AUTHOR CONTRIBUTIONS

Y. K. and Z. L. contributed equally to this work.

## ACKNOWLEDGMENTS

This work was supported by the National Natural Science Foundation of China (Grants No. 92265113, No. 12074368, and No. 12034018), by the Innovation Program for Quantum Science and Technology (Grant No. 2021ZD0302300). This work was partially carried out at the USTC Center for Micro- and Nanoscale Research and Fabrication.